\documentclass[aps,floatfix,nofootinbib,showpacs,twocolumn,superscriptaddress]{revtex4} 
 
\usepackage{amsmath} 
\usepackage{dcolumn} 
\usepackage{graphicx} 

\newcommand{\sfrac}[2]{\mbox{\footnotesize $\displaystyle \frac{#1}{#2}$}}

\begin{document} 
 
 
\title{Perspective on rainbow-ladder truncation}
 
\author{G.~Eichmann}
\affiliation{Physics Division, Argonne National Laboratory, 
             Argonne IL 60439-4843 U.S.A.} 
\affiliation{Institut f\"ur Physik, Karl-Franzens-Universit\"at Graz, A-8010 Graz, Austria}
 
\author{R.~Alkofer}
\affiliation{Institut f\"ur Physik, Karl-Franzens-Universit\"at Graz, A-8010 Graz, Austria}

\author{I.\,C.~Clo\"et} 
\affiliation{Physics Division, Argonne National Laboratory, 
             Argonne IL 60439-4843 U.S.A.} 

\author{A.~Krassnigg}
\affiliation{Institut f\"ur Physik, Karl-Franzens-Universit\"at Graz, A-8010 Graz, Austria}
                          
\author{C.\,D.~Roberts} 
\affiliation{Physics Division, Argonne National Laboratory, 
             Argonne IL 60439-4843 U.S.A.} 
             
\begin{abstract} 
\rule{0ex}{3ex} 
\emph{Prima facie} the systematic implementation of corrections to the rainbow-ladder truncation of QCD's Dyson-Schwinger equations will uniformly reduce in magnitude those calculated mass-dimensioned results for pseudoscalar and vector meson properties that are not tightly constrained by symmetries.  The aim and interpretation of studies employing rainbow-ladder truncation are reconsidered in this light.
\end{abstract} 
 
\maketitle 
 


That the lightest bound states supported by QCD are members of a $J^{PC}=0^{-+}$ isotriplet owes to the near chiral symmetry of the light-quark Lagrangian and the pattern by which it is dynamically broken.  The particular behaviour of the pion mass; viz., $m_\pi^2 \propto \hat m$, where $\hat m$ is the renormalisation-group-invariant light-quark current-mass, and numerous other fascinating properties can be derived in QCD through a careful consideration of the axial-vector Ward-Takahashi identity and the gap equation; e.g.,  Ref.\,\cite{Maris:1997hd,Bicudo:2003fp,Holl:2004fr,Holl:2005vu,Bhagwat:2007ha}.  In relation to the latter, Dyson-Schwinger equation studies \cite{Bhagwat:2003vw,Alkofer:2003jj} and simulations of the lattice-regularised theory \cite{Bowman:2005vx} have established conclusively that dynamical chiral symmetry breaking (DCSB) is a reality in QCD.  These facts have been used to provide a satisfactory understanding of the pion's mass.

Such is not the case, e.g., for the pion's radius nor its weak decay constant.  They are both well known experimentally \cite{Yao:2006px}: $r_\pi = 0.672 \pm 0.008\,$fm; $f_\pi=130.7 \pm 0.4\,$GeV, but the problem of understanding them within QCD remains.  Their reliable calculation requires more detailed knowledge of strong interaction dynamics than does $m_\pi$.  Furthermore, one must employ a formulation of the quantum field theory bound state problem that veraciously incorporates DCSB, and its corollaries, and simultaneously preserves the vector Ward-Takahashi identity.

This can be achieved in a Poincar\'e covariant and symmetry preserving treatment of quark-antiquark bound states through the Bethe-Salpeter equation (BSE)
\begin{equation}
\label{bse1}
\Gamma_{tu}(k;P) = \int^\Lambda_q [\chi(q;P)]_{sr}\, K_{rs}^{tu}(q,k;P)\,,
\end{equation}
where: $k$ is the relative and $P$ the total momentum of the constituents; $r$,\ldots,\,$u$ represent colour, Dirac and flavour indices; $\chi(q;P):= S(q_+) \Gamma(q;P) S(q_-)$, with $\Gamma(q;P)$ a given meson's Bethe-Salpeter amplitude and $q_\pm = q\pm P/2$; and $\int^\Lambda_q$ represents a translationally invariant regularisation of the integral, with $\Lambda$ the regularisation mass-scale \cite{Maris:1997hd,Maris:1997tm}.  In Eq.\,(\ref{bse1}), $S$ is the renormalised dressed-quark propagator and $K$ is the fully amputated dressed-quark-antiquark scattering kernel.  The product $SS K$ is a renormalisation point invariant.  

The dressed-quark propagator appearing in the BSE's kernel is determined by the renormalised gap equation\footnote{In our Euclidean metric:  $\{\gamma_\mu,\gamma_\nu\} = 2\delta_{\mu\nu}$; $\gamma_\mu^\dagger = \gamma_\mu$; $\gamma_5= \gamma_4\gamma_1\gamma_2\gamma_3$; $a \cdot b = \sum_{i=1}^4 a_i b_i$; and $P_\mu$ timelike $\Rightarrow$ $P^2<0$.}
\begin{eqnarray}
S(p)^{-1} & =&  Z_2 \,(i\gamma\cdot p + m^{\rm bm}) + \Sigma(p)\,, \label{gendse} \\
\Sigma(p) & = & Z_1 \int^\Lambda_q\! g^2 D_{\mu\nu}(p-q) \frac{\lambda^a}{2}\gamma_\mu S(q) \Gamma^a_\nu(q,p) , \label{gensigma}
\end{eqnarray}
where $D_{\mu\nu}(k)$ is the dressed-gluon propagator, $\Gamma_\nu(q,p)$ is the dressed-quark-gluon vertex, and $m^{\rm bm}$ is the $\Lambda$-dependent current-quark bare mass.  The quark-gluon-vertex and quark wave function renormalisation constants, $Z_{1,2}(\zeta^2,\Lambda^2)$, depend on the renormalisation point, $\zeta$, the regularisation mass-scale and the gauge parameter.  The gap equation's solution has the form 
\begin{eqnarray} 
 S(p)^{-1} & = & i \gamma\cdot p \, A(p^2,\zeta^2) + B(p^2,\zeta^2) \,.
%
\label{sinvp} 
\end{eqnarray}
and the mass function $M(p^2)=B(p^2,\zeta^2)/A(p^2,\zeta^2)$ is renormalisation point independent.  The propagator is obtained from Eq.\,(\ref{gendse}) augmented by the condition
\begin{equation}
\label{renormS} \left.S(p)^{-1}\right|_{p^2=\zeta^2} = i\gamma\cdot p + m(\zeta)\,,
\end{equation}
where $m(\zeta)$ is the renormalised mass: 
$Z_2(\zeta^2,\Lambda^2) \, m^{\rm bm}(\Lambda)$ $=$ $Z_4(\zeta^2,\Lambda^2) \, m(\zeta)$,
with $Z_4$ the Lagrangian mass renormalisation constant.  The chiral limit means $\hat m = 0$.

The Bethe-Salpeter and gap equations are Dyson-Schwinger equations (DSEs), for which a nonperturbative Poincar\'e covariant and symmetry preserving truncation scheme is available \cite{Munczek:1994zz,Bender:1996bb}.  The leading-order term is the so-called rainbow-ladder truncation.  
%
%
We assume it is an accurate tool for calculating meson properties in the heavy-quark limit because for relevant momenta the contribution from additional diagrams diminishes as the current-quark mass of a bound state's constituents increases \cite{Bhagwat:2004hn}.
Furthermore, in equal-mass-constituent pseudoscalar and vector meson channels it can be seen algebraically that contributions from subleading diagrams interfere destructively amongst themselves owing to the axial-vector Ward-Takahashi identity \cite{Bender:1996bb,Bhagwat:2004hn,Matevosyan:2006bk}.

These facts have successfully been exploited through the application of a renormalisation-group-improved rainbow-ladder DSE model \cite{Maris:1997tm,maristandy1} to a wide variety of phenomena involving $\pi$- and $\rho$-mesons \cite{revpieter}.  The heart of the model is an \textit{Ansatz} for the Bethe-Salpeter kernel:
\begin{eqnarray}
\nonumber \lefteqn{
K^{tu}_{rs}(q,k;P) = - \,{\cal G}((k-q)^2) }\\
&&  \times \, D_{\mu\nu}^{\rm free}(k-q)\,\left[\gamma_\mu \frac{\lambda^a}{2}\right]_{ts} \, \left[\gamma_\nu \frac{\lambda^a}{2}\right]_{ru} \!, \label{ladderK}
\end{eqnarray}
wherein $D_{\mu\nu}^{\rm free}(\ell)$ is the free gauge boson propagator and \cite{Bloch:2002eq,Maris:2002mt,Eichmann:2007nn}
\begin{eqnarray}
\nonumber \lefteqn{
\frac{1}{Z_2^2}\frac{{\cal G}(s)}{s} =  {\cal C}(\omega,\hat m) \, \frac{4\pi^2}{\omega^7} \,\frac{s}{\Lambda_t^4}\, {\rm e}^{-s/[\omega^2 \Lambda_t^2]}}\\
&& + \frac{8\pi^2 \gamma_m}{\ln\left[ \tau + \left(1+s/\Lambda_{\rm QCD}^2\right)^2\right]} \, {\cal F}(s)\,,
\label{Gkmodel}
\end{eqnarray}
with ${\cal F}(s)= [1-\exp(-s/\Lambda_t^2)]/s$, $\Lambda_t=1.0\,$GeV, $\tau={\rm e}^2-1$, $\gamma_m=12/25$ and $\Lambda_{\rm QCD} = \Lambda^{(4)}_{\overline{MS}} = 0.234\,$GeV.  

This form expresses the interaction as a sum of two terms.  The second ensures that perturbative behaviour is correctly realised at short range; namely, as written, for $(k-q)^2 \sim k^2 \sim q^2 \gtrsim 1 - 2\,$GeV$^2$, $K$ is precisely as prescribed by QCD.  On the other hand, the first term in ${\cal G}(k^2)$ is a model for the long-range behaviour of the interaction.  It is a finite width representation of the form discussed in Refs.\,\cite{mn83}.

Given a truncation of the BSE's kernel there is a unique gap equation that ensures the Ward-Takahashi identities are automatically satisfied \cite{Munczek:1994zz,Bender:1996bb,Bhagwat:2004hn,Matevosyan:2006bk}.  In connection with Eq.\,(\ref{ladderK}) that is a rainbow gap equation; viz., Eq.\,(\ref{gendse}) with \begin{equation} 
%
\Sigma(p)=\int^\Lambda_q\! {\cal G}((p-q)^2) D_{\mu\nu}^{\rm free}(p-q) \frac{\lambda^a}{2}\gamma_\mu S(q) \frac{\lambda^a}{2}\gamma_\nu . \label{rainbowdse} 
\end{equation} 

Having selected a DSE truncation and specified the behaviour of the interaction, one can straightforwardly proceed to calculate hadron properties.  There are apparently two active parametric quantities in Eq.\,(\ref{Gkmodel}); viz., ${\cal C}$ and $\omega$, which together determine the integrated infrared strength of the rainbow-ladder kernel.  They were fitted to a selection of physical observables in Ref.\,\cite{maristandy1}.  However, in reconsidering that fit it was noted \cite{Maris:2002mt} that the predictions are approximately independent of $\omega$ on the domain $\omega\in[0.3,0.5]$.  Hence, there is truly only one parameter and experimental values are satisfactorily reproduced with ${\cal C}(\omega,\hat m)={\rm const.}=0.37$.  This observation has since been much exploited.  Nevertheless, we choose to reconsider this approach and explore whether there is a more appropriate way to determine ${\cal C}(\omega,\hat m)$. 

\begin{figure}[t]
\centerline{
\includegraphics[clip,width=0.43\textwidth]{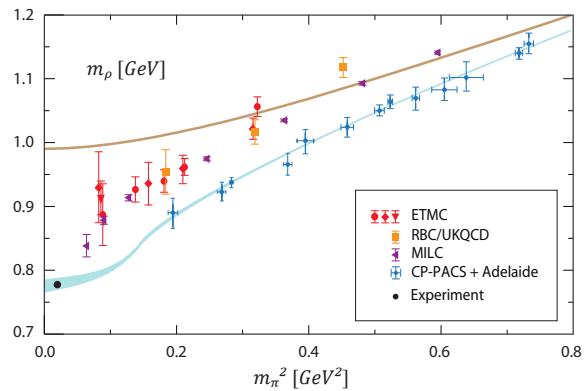}}
\caption{\label{fig1} \emph{Solid line}: expected evolution in rainbow-ladder truncation of the $\rho$-meson mass with pion mass-squared, Eq.\,(\protect\ref{masstrajectory}).  Results from simulations of lattice-regularised QCD are also depicted: ETMC \cite{Boucaud:2007uk}; RBC/UKQCD \cite{Allton:2007hx}; and MILC \cite{Bernard:2001av}, plus the results in Ref.\,\protect\cite{AliKhan:2001tx} with an analysis and chiral extrapolation \protect\cite{Allton:2005fb}.  In studying this figure it is important to bear in mind that Eq.\,(\protect\ref{masstrajectory}) is a calculated overestimate of $m_\rho(m_\pi^2)$.} 
\end{figure}

To develop an answer we remark that in pseudoscalar and vector channels the net effect of corrections to the rainbow-ladder truncation is attraction.  Hence at the physical light-quark current-mass the truncation should yield a value for $m_\rho$ that is larger than experiment.  Extant studies indicate that a reasonable value is $m_\rho^0:=m_\rho^{\rm RL}(\hat m=0) = 0.99\,$GeV because the resummation of nonresonant diagrams \cite{Bhagwat:2004hn,Matevosyan:2006bk,Roberts:2007jh} produces $\sim 100\,$MeV of attraction and pseudoscalar meson loops provide a further $\sim 120\,$MeV, in addition to generating a width \cite{Pichowsky:1999mu}.  Corrections to rainbow-ladder vanish with increasing current-quark mass.  For example, at $\hat m$ such that the mass of a pion-like pseudoscalar is $0.63\,$GeV, corrections to $m_\rho$ from meson loops are already negligible.  Furthermore, nonresonant contributions are $70$\% of their chiral-limit magnitude; at $m_\pi=0.9\,$GeV that drops to $50$\%; and they vanish completely in the heavy-quark limit.  We express this behaviour through (see Fig.\,\ref{fig1}):
\begin{equation}
\label{masstrajectory}
\frac{m_\rho^2(m_\pi^2)}{(m_\rho^0)^2} = 1 + \frac{(m_\pi/m_\rho^0)^4}{0.6 + (m_\pi/m_\rho^0)^2}\,.
\end{equation}

\begin{figure}[t]
\centerline{\includegraphics[clip,width=0.47\textwidth]{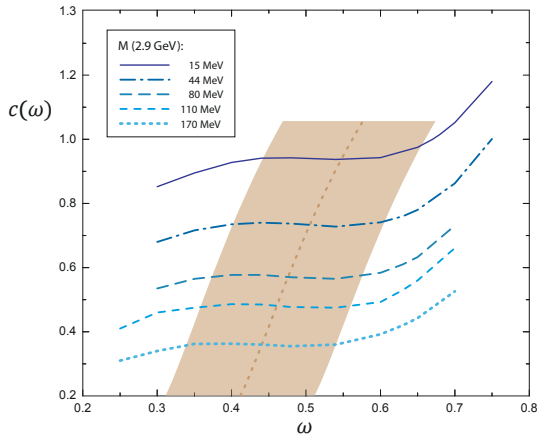}}
\caption{\label{fig2} ${\cal C}(\omega,\hat m)$ in Eq.\,(\ref{Gkmodel}) determined by Eq.\,(\protect\ref{masstrajectory}).  Five current-quark masses are shown explicitly.  The shaded region demarcates a domain of minimal sensitivity to this model parameter.  The dashed almost-vertical curve indicates $\bar\omega(\hat m )=0.38+0.17/[1+ \hat m/\hat m_0]$, $\hat m_0=0.12\,$GeV.} 
\end{figure}

Since vector meson masses are observable, well known and not constrained by symmetries, we determine ${\cal C}(\omega,\hat m)$ by requiring that, with the interaction model defined in Eq.\,(\ref{Gkmodel}), rainbow-ladder truncation reproduce Eq.\,(\ref{masstrajectory}).  The procedure is straightforward.  For a given current-quark mass, and $\omega$ in Eq.\,(\ref{Gkmodel}), the rainbow-ladder truncation produces values for $m_\rho$ and $m_\pi$ via Eqs.\,(\ref{bse1}), (\ref{gendse}), (\ref{ladderK}) and (\ref{rainbowdse}).  ${\cal C}(\omega,\hat m)$ is chosen such that these values are related through Eq.\,(\ref{masstrajectory}).  The result is depicted in Fig.\,\ref{fig2}, which also illustrates clearly the observation of Ref.\,\cite{Maris:2002mt}; viz., for each $\hat m$ there is a domain of $\omega$ on which Eq.\,(\ref{masstrajectory}) is preserved with ${\cal C}(\omega,\hat m)\approx\,$constant.  The shaded band in Fig.\,\ref{fig2} highlights that domain.  We will subsequently take this band to define a domain of internally consistent interaction tensions \cite{Hecht:2000ga} and report physical quantities for all values of $\omega$ within this domain.  The variation in results should be taken as an indication of the model-dependent uncertainty in our predictions.  An accurate interpolation within this band is 
\begin{equation}
\label{Comegam}
\begin{array}{l}
\displaystyle {\cal C}(\omega,\hat m) ={\cal C}_0 +
\frac{{\cal C}_1(\omega-\bar\omega(\bar m))}{1 + {\cal C}_2 \,\hat x + {\cal C}_3^2\, \hat x^2},\;\hat x = \hat m / \hat m_0,\\[2.5ex]
{\cal C}_1(\ell) = 0.86 ( 1 - 0.15 \ell  + (1.50 \ell)^2 + (2.95\ell)^3)\,,
\end{array}
\end{equation}
with $\bar\omega(\hat m)$ the $\hat m$-dependent midpoint of the shaded domain in Fig.\,\ref{fig2} and ${\cal C}_0=0.11$, ${\cal C}_2=0.885$, ${\cal C}_3=0.474$.  

\begin{figure}[t]
\centerline{\includegraphics[width=0.43\textwidth]{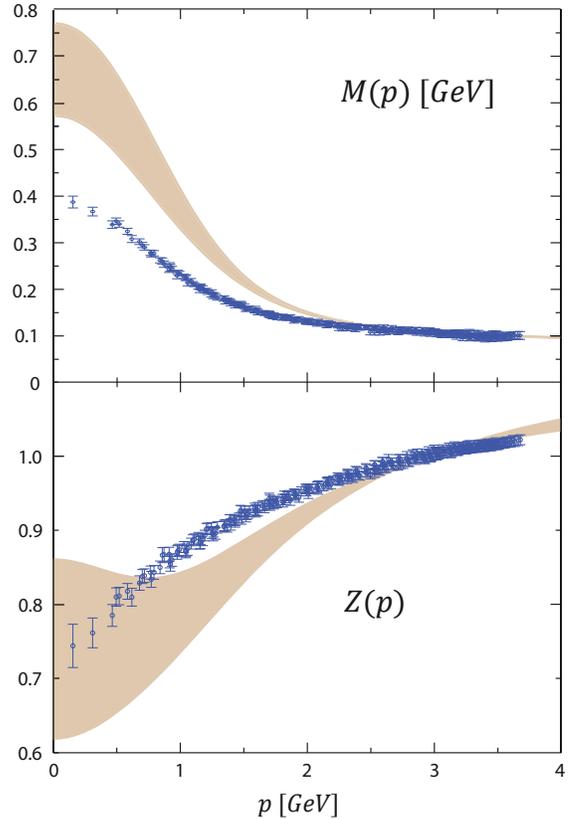}}
\caption{\label{plotprop} \emph{Top panel} -- mass function and \emph{bottom panel} -- wave function renormalisation, $Z(p,\zeta)=1/A(p,\zeta)$, calculated with $\zeta=2.9\,$GeV and $\hat m$ such that $m_\pi=0.63(1)\,$GeV.  For $M(p^2)$, the lower (upper) edge of the bands corresponds to the largest (smallest) $\omega$ value in the domain demarcated in Fig.\,\protect\ref{fig2} -- this pattern is reversed for $Z(p,\zeta)$.  Circles in each panel are results from simulations of lattice-regularised QCD \protect\cite{Bowman:2005vx}.}
\end{figure}

Figure~\ref{plotprop} illustrates the impact of this analysis on the behaviour of the functions in Eq.\,(\ref{sinvp}), which characterise the dressed-quark propagator.  
For relevant momenta, corrections to the rainbow truncated gap equation are suppressed by increasing $\hat m$.
%
%
Furthermore, while they are always negligible for $p^2\gtrsim 3\,$GeV$^2$, corrections suppress $M(p^2)$ and $Z(p^2)$ for $p^2 \lesssim 1\,$GeV$^2$, while maintaining their support at intermediate momenta \cite{Bhagwat:2004hn}.  

We have now determined the dressed-quark propagator and meson Bethe-Salpeter amplitudes, and their current-quark mass dependence.  Physical quantities can therefore be calculated; e.g., the vacuum condensate \cite{Maris:1997hd}, 
\begin{equation}
\label{qbq0} \,-\,\langle \bar q q \rangle_\zeta^0 = \lim_{\Lambda\to \infty} 
Z_4(\zeta^2,\Lambda^2)\, N_c \, {\rm tr}_{\rm D}\int^\Lambda_q\!
S^{0}(q,\zeta)\,,  \\
\end{equation}
where the superscript ``0'' denotes a quantity calculated in the chiral limit, and the $\pi$ and $\rho$ decay constants \cite{Maris:1997hd,Ivanov:1998ms}:
\begin{eqnarray}
\label{fpin} f_{\pi^+}  \,  P_\mu &=& Z_2\,{\rm tr}_{\rm CD} \int^\Lambda_q \gamma_5\gamma_\mu\, \chi^{\pi^+}(q;P) \,, \\
f_{\rho^+} m_\rho &=& \sfrac{1}{3} Z_2\,{\rm tr}_{\rm CD}\int^\Lambda_q \gamma_\mu\, \chi_{\mu}^{\rho^+}(q;P)\,.
\end{eqnarray}

\begin{table}[t] 
\caption{\label{tableresults} \emph{Row 1}: Results calculated with $\hat m = 6.1\,$MeV, for which $m_\pi = 0.138\,$GeV.  Error describes sensitivity to variation in $\omega$ around $\bar\omega$ in Fig.\,\protect\ref{fig2}.  \emph{Row 2}: Experimental \protect\cite{Yao:2006px} or phenomenological values. All quantities in GeV, except $r_\pi$ in fm.  The condensate is quoted at $1\,$GeV after one-loop evolution from the renormalisation scale $\zeta=19\,$GeV, as described in Ref.\,\protect\cite{Maris:1997tm}. The axial-vector Ward-Takahashi identity ensures that subleading corrections to the pion's mass are $\lesssim 5$\% \protect\cite{Bhagwat:2004hn}.}
\begin{ruledtabular} 
\begin{tabular*} 
{\hsize} {l@{\extracolsep{0ptplus1fil}} |
l@{\extracolsep{0ptplus1fil}}
l@{\extracolsep{0ptplus1fil}}
l@{\extracolsep{0ptplus1fil}}
l@{\extracolsep{0ptplus1fil}}
l@{\extracolsep{0ptplus1fil}}} 
 & $[-\langle \bar q q \rangle_1^0]^{\frac{1}{3}}$ &  $f_{\pi^+}$ & $f_{\rho^+}$ & $m_{\rho}$ & $r_\pi$\\\hline
Calc. & 0.319(2) & 0.178 & 0.283(9) & 0.99 & 0.485(3)\\
Phen. & 0.236  & 0.131 & 0.216 & 0.77 & 0.672
\end{tabular*} 
\end{ruledtabular} 
\end{table} 

In Table~\ref{tableresults} we present results obtained in rainbow-ladder truncation using Eqs.\,(\ref{Gkmodel}) and (\ref{Comegam}) at a current-quark mass for which $m_\pi=0.138\,$GeV.  Expressed in units of mass, the tabulated quantities overestimate experiment by $34 \pm 4\,$\%.  This uniform response to a deliberate increase in the interaction tension is internally consistent; viz., a single cause produces a uniform effect and that permits a single remedy.  NB.\  Attempts to estimate the contribution from pseudoscalar meson loops alone produced uncorrected values of $[-\langle \bar q q \rangle_1^0]^{\frac{1}{3}} \sim 0.28\,$GeV and $f_{\pi^+} \sim 0.17\,$GeV \cite{Blaschke:1995gr,Oertel:2000jp}, comparable to Row~1.

\begin{figure}[t]
\centerline{\includegraphics[width=0.45\textwidth]{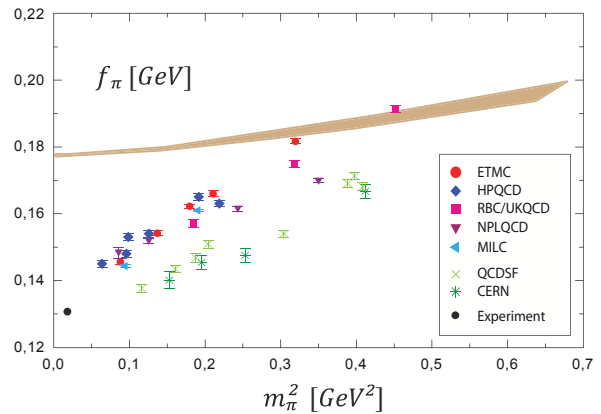}}
\caption{\label{fig3} \emph{Shaded band}: calculated mass dependence of the pion's leptonic decay constant.  The band delimits the range of results possible as $\omega$ varies over the domain in Fig.\,\protect\ref{fig2}.  Results from simulations of lattice-regularised QCD are also presented: ETMC \cite{Boucaud:2007uk}; RBC/UKQCD \cite{Allton:2007hx}; HPQCD \cite{Follana:2007uv}; NPLQCD \cite{Beane:2006kx}; MILC \protect\cite{Bernard:2007ps}; QCDSF \cite{Gockeler:2006ns}; and CERN \cite{DelDebbio:2006cn,DelDebbio:2007pz}, a compilation drawn from Ref.\,\cite{McNeile:2007fu}.}
\end{figure}

In Fig.\,\ref{fig3} we depict the calculated current-quark mass dependence of the pion's leptonic decay constant.  A material similarity with Fig.\,\ref{fig1} is apparent.  NB.\ The discussion associated with Eq.\,(\ref{masstrajectory}) should be borne in mind when contemplating this figure.

In rainbow-ladder truncation the electromagnetic pion form factor is determined from \cite{Roberts:1994hh}
\begin{eqnarray}
\nonumber
\lefteqn{(p_1+p_2)_\mu F_\pi(Q^2) = 
\text{tr}_{\rm CD} \int_k \bar {\Gamma}_{\pi^+}(k;-p_2) \, S(k_{+}^{+}) }\\
&\times &  \Gamma_{\pi^+}(k_{0}^{+};p_1) \,  S(k_{-}^{+}) \, i \Gamma_\mu(k_{-}^{+},k_{-}^{-}) \, S(k_{-}^{-})\,
 , \label{current:em}
\end{eqnarray}
where $k_{\alpha}^{\beta}=k+\alpha p_1/2+\beta Q/2$ and $p_2=p_1+Q$.  The only quantity in Eq.\,(\ref{current:em}) that we have not yet encountered is the dressed-quark photon vertex, $\Gamma_\mu(\ell_1,\ell_2)$.  It is obtained from an inhomogeneous Bethe-Salpeter equation and, in a symmetry preserving DSE truncation, the solution satisfies the vector Ward-Takahashi identity.  

The definitive rainbow-ladder calculation of $F_\pi(Q^2)$ solved the vertex equation and employed the result in Eq.\,(\ref{current:em}) \cite{Maris:1999bh}.  It can be argued from this study that a reliable approximation to the result for the pion's charge radius in rainbow-ladder truncation is provided by \cite{Maris:1999bh,Roberts:2000aa}
\begin{equation}
\label{piradius}
r_\pi^2 = r_{\pi\, BC}^2 + \frac{6}{m_\rho^2} \frac{g_{\rho \pi \pi}(0)}{g_\rho} \,{\rm e}^{-\varrho}.
\end{equation}

The first term in Eq.\,(\ref{piradius}), $r_{\pi\, BC}$, is the nonresonant contribution to the radius.  It is obtained from Eq.\,(\ref{current:em}) with the Ball-Chiu \textit{Ansatz} for $\Gamma_\mu(\ell_1,\ell_2)$ \cite{Ball:1980ay,Roberts:1994hh}.
%
%
%
The second term represents the $\rho$-meson contribution to the radius.  That is explicitly excluded from Eq.\,(\ref{current:em}) when one employs the Ball-Chiu vertex calculated from the solution of the rainbow gap equation.  The quantity $g_\rho = \surd 2 m_\rho/f_\rho$ is the $\rho^0 \to e^+ e^-$ coupling.  The function $g_{\rho \pi \pi}(Q^2)$ describes the $\rho\pi\pi$ vertex, which in rainbow-ladder truncation is given by a straightforward analogue of Eq.\,(\ref{current:em}).  Experimentally, $g_{\rho\pi\pi}(Q^2=-m_\rho^2)=6.14$ but Eq.\,(\ref{piradius}) requires an off-shell-$\rho$ coupling, $g_{\rho\pi\pi}(Q^2=0)$.  We calculate that with the on-shell $\rho$-meson Bethe-Salpeter amplitude, modified such that each subleading Dirac structure acquires a suppression factor: $[2+Q^2/m_\rho^2]^{-1/2}$.  The factor ${\rm e}^{-\varrho}$ in Eq.\,(\ref{piradius}) accounts for any additional effect owing to an off-shell $\rho$-meson.\footnote{A single-pole vector meson dominance \textit{Ansatz} assumes $g_{\rho\pi\pi}(Q^2=0)=g_\rho$, $\varrho=0$.  The latter is also assumed in Ref.\,\protect\cite{Maris:1999bh}.}  Its value is determined by requiring that Eq.\,(\ref{piradius}) reproduce the DSE curve in Fig.\,6 of Ref.\,\cite{Maris:2005tt}, with the result $\varrho = 0.2\,m_\pi$.  This analysis gives $g_{\rho\pi\pi}(0){\rm e}^{-\varrho} \approx\,$ constant$\,=2.75(40)$ on the relevant $(\omega,m_\pi)$ domain: larger $\omega$, smaller constant value, and vice-versa. 

\begin{figure}[t]
\centerline{\includegraphics[width=0.45\textwidth]{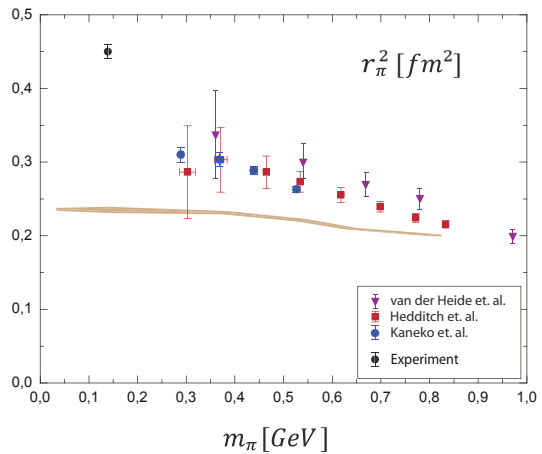}}
\caption{\label{fig4} \emph{Shaded band}: calculated pion-mass dependence of the pion's mean-square charge radius.  The $\omega$-variation band is narrow owing to cancellations between effects on the two terms in Eq.\,(\protect\ref{piradius}).  Results from simulations of lattice-regularised QCD are also presented: quenched, triangles and squares \cite{vanderHeide:2003kh,Hedditch:2007ex}; and unquenched, circles \cite{Kaneko:2007nf}.}
\end{figure}

Thus specified, Eq.\,(\ref{piradius}) yields the charge radius in Table\,\ref{tableresults} and the evolution with $m_\pi$ depicted in Fig.\,\ref{fig4}.  On the material $(\omega,m_\pi)$ domain the nonresonant contribution provides approximately 50\% of the reported value of $r_\pi^2$.  Furthermore, on this domain we calculate $r_\pi f_\pi = 0.44(1)$; viz., a constant independent of the current-quark mass.  The experimental value is $0.445\pm0.007$.  As could be anticipated from Ref.\,\cite{Maris:1998hc}, with $r_{\rm BC}$ alone the calculated value is only $\sim 0.32$.  It is noteworthy that an estimate of the contribution to the charge radius owing to pseudoscalar meson loops gives \cite{Alkofer:1993gu} $r_{\pi\,{\rm loop}}^2 \approx 0.07(2)\,$fm$^2$ at $m_\pi=0.14\,$GeV.  This must simply be added to our result because the $\rho$-meson term in Eq.\,(\ref{piradius}) has zero width.  The result is $r_\pi = 0.55(2)$.  Naturally, as $m_\pi$ increases the $\pi$-loop contribution diminishes rapidly in importance.

We have explained that in connection with light-quark systems, and those of the physical qualities of the pseudoscalar and vector meson bound states they constitute which are not tightly constrained by symmetries, the rainbow-ladder truncation of QCD's DSEs should produce results that, when measured in units of mass, are uniformly $\approx 35$\% too large.  The systematic implementation of corrections will then shift calculated results so that reliable predictions and agreement with experiment can subsequently be expected.  In this way one can arrive at a veracious understanding of light-quark observables.

Our study also explains why, when employed as a means of modelling QCD, it is possible to tune the parametric elements in a rainbow-ladder truncation such that a wide range of light-quark pseudoscalar- and vector-meson observables can successfully be correlated \cite{revpieter}.  That rainbow-ladder is the first term in a systematic truncation guarantees a uniform response to corrections and hence allows for their effects in pseudoscalar and vector channels to be expressed in large part through simple parameter modifications.  Naturally, however, features such as decay widths and associated nonanalyticities, which are essentially tied to hadron loops, cannot be realised in this way.

%
We acknowledge interactions with B.~El-Bennich, T.~Kl\"ahn and R.\,D.~Young.
This work was supported by: the Department of Energy, Office of Nuclear Physics, contract no.\ DE-AC02-06CH11357; \emph{Deutsche Forschungsgemeinschaft} under grant no.\ Al279/5-1 \& 5-2; and the Austrian Science Fund \emph{FWF} under grant no.\ W1203 
and project no.\ P20496-N16.
%

\end{document}